\documentstyle[proceedings,numreferences,epsf]{crckapb}

\begin{opening}
\title{Gravitational Waves Generated by Globular\break
Cluster Systems Collapse}

\author{R. Capuzzo -- Dolcetta}
\institute{Istituto Astronomico (Univ. of Rome ``La Sapienza'')\\
           via G.M. Lancisi 29, I-00161 - Rome, Italy}
\author{P. Miocchi}
\institute{Dep. of Physics (Univ. of Rome ``La Sapienza'')\\
           P.le Aldo Moro 5, I-00185 - Rome, Italy}
\end{opening}

\runningtitle{Gravitational Waves from Galactic Black Hole}
\begin{document}
The hypothesis --raised and discussed in \cite{Cap1}--  that the
source of AGN fueling can be, at least in elliptical galaxies, stars belonging
to a dense stellar environment formed by globular clusters frictionally
decayed and tidally destroyed by the gravitational field of central black
hole (b.h.), is strongly supported in \cite{Cap6}.
The quantity of star mass swallowed by the b.h. and its time rate seem at all
compatible with those needed to justify the intensity and time characteristics
of galactic nuclei activity  (see also \cite{Cap2}, \cite{Cap3}, \cite{Cap4}).

In this short report we give a preliminary estimate of the number of
gravitational impulses per year that the orbiting gravitational antenna LISA
(at present under development by ESA) should detect. They
are emitted by the stars falling into the b.h. whose mass, $M(t)$,
and accreting rate per unit mass, $\dot m$, are given by the mentioned model
(see \cite{Cap6}, \cite{Cap3}).

We make the following assumptions: i) all the stars fall {\it radially} into
the b.h.; ii) all such stars have $R=R_\odot$, and $m=M_\odot$. We
know that a point mass $m$ in a radial free fall into a b.h. of mass $M$,
emits gravitational waves in form of an impulse with a well defined amplitude
and spectrum (see \cite{davis}).

Moreover if a star has a radius very small compared with the
wavelength of the gravitational waves emitted, then this star can be treated
as a point mass. Hence destructive interference phenomena can be neglected
(see \cite{haugan})
and the amplitude of the emitted impulse is $A\sim 0.49(Gm/c^2)/d$ where $d$
is the distance between the b.h. and the observer.
Furthermore the detector LISA has the lowest amplitude limit
$A_{min}\sim 10^{-23}$ in the frequency range $10^{-3}\div 10^{-2}\mbox{Hz}$.
Consequently a source has to satisfy some conditions to be detected:
i) it must be close enough for the impulse amplitude be $A>A_{min}$;
ii) the spectrum of the impulse has to be peaked inside the frequency range
above; iii) there must be no destructive interferences.
These conditions involve the b.h. mass $M$ and its distance
(see \cite{Cap6} and \cite{Cap5} for more details).

\begin{figure}[htb]
\vskip 1.2 truecm
\epsfysize 4.5 true cm
\epsffile{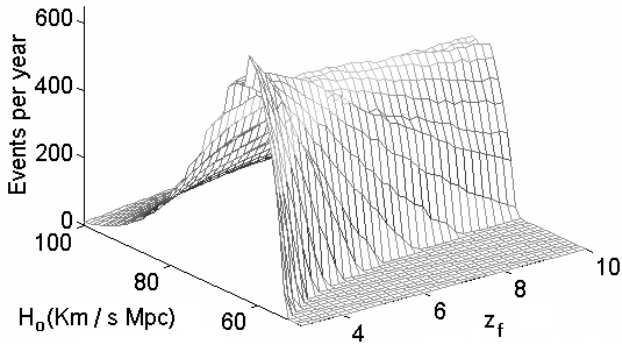}
\caption{Number of impulses per year that LISA should detect as a function
of $H_0$ and of the red-shift of galaxy formation, $z_f$.\label{res}}
\end{figure}

Thus, considering {\it all} possible sources, integrating
their contribution over all red-shifts between $z=0$ and $z=\hat z$, $\hat z$
being the red-shift corresponding to the maximum distance below which
$A>A_{min}$, we have the number of impulses per year that LISA should
detect:
\begin{equation}
N=\int_0^{\hat z} \Omega(z)n(z)\Gamma(z)dz. \label{int}
\end{equation}
In the integral (\ref{int}), $n(z)$ is the density of galaxies at red-shift
$z$ that we presume they have a central b.h., $\Gamma(z)dz$ is the volume
element and $\Omega(z)$ is equal to $\dot m$ if the conditions above are
satisfied (null otherwise).
Considering $z=(2/3H_0t)^{2/3}-1$ (flat universe) and
adopting $n(0)=0.1 \mbox{\ Mpc}^{-3}$ we obtain the result reported
in fig.~\ref{res} in which the number of impulses detected
per year depends on the value of the Hubble constant and on the red-shift of
galaxy formation $z_f$. This latter represents the origin of time by which
our AGN fueling hypothesis starts to work.
\\

\end{document}